# The Signal Space Separation method

Samu Taulu, Matti Kajola and Juha Simola

Elekta Neuromag Oy, Helsinki, Finland
**Corresponding author**: Samu Taulu, Elekta Neuromag Oy, Elimäenkatu 22 A, 00510 Helsinki, Finland.
Email: samu.taulu@neuromag.fi

## Introduction

Extraction of the weak biomagnetic signals from multichannel measurements dominated by environmental interference sources is a basic problem in biomagnetic recordings. Traditional methods to attack this problem include magnetically shielded rooms, gradiometers, and software compensation, such as the signal space projection (SSP) [1], or extrapolation based on reference channels [2]. SSP is capable of reducing the interference signals by a factor of several thousands. It also changes the appearance of the biomagnetic signals though this change is usually small and easy to take into account in the modeling. On the other hand, the reference channel method assumes that any interference seen by the signal channels can be modeled and compensated for by a small number of reference channels. In order to be sensitive to the interference only, the reference channels need to be located far away from the signal sensors, which increases the extrapolation distance and limits the capability of removing signals from nearby interference sources.

Signal Space Separation (SSS) [3] is a new method for compensation of external interference and sensor artifacts. This method is based on the fact that modern multichannel MEG devices with the number of signal channels exceeding 300 provide generous oversampling of the magnetic fields consisting of both biomagnetic and interference fields. The oversampling condition is true for all fields arising from sources located about or more than two centimeters away from the nearest sensor in the array. The magnetic fields produced by such sources form the set of possible magnetic signals, the magnetic subspace, having a dimensionality less than about 150 in practical measurements [4].

The relatively low dimension of the magnetic subspace is based on the fact that all sensors are located in a volume free of sources of the magnetic field. Thus, the magnetic field must be derivable from a harmonic scalar potential which is spatially quite a smooth function because of the distance between the sensors and the sources. The harmonic scalar potential can be represented as a truncated expansion of harmonic basis functions by leaving out the high-order terms representing unmeasurable fine details. The signal vectors obtained by evaluating the harmonic basis functions for all channels span a signal space containing all signals related to magnetic fields.

The crucial observation in the SSS method is that there are separate basis functions for signals arising from inside of the sensor array and for signals arising from the environment surrounding the sensor array. Consequently, the magnetic subspace contains two linearly independent subspaces: one for the interesting biomagnetic signals and one for the interference signals.

In this way a unique decomposition of the measured signal vector with separate components for the interesting and interference signals is obtained. The decomposition is based on proper signal channels with no need for dedicated reference channels. The interesting signal can be reconstructed from the components corresponding to the interesting subspace. As a consequence, the morphology and amplitude of the interesting signal do not change.

The decomposition of the magnetic field is device-independent when evaluated in a coordinate system attached to the subject's head. Thus, the decomposition can be used to transform signals between different sensor configurations. This also enables an effective movement correction method comprising of disturbance removal followed by a virtual signal calculation for the desired reference position of the head. The movement correction method also leads to another interesting application: it can be used to separate and model magnetic fields arising from DC sources.

## Methods

The sensors of the multichannel devices used in biomagnetic recordings are located in a source-free volume. Thus, the magnetic field in that volume is a gradient of a harmonic scalar potential $V$:

$$\mathbf{b} = -\nabla V, \quad \nabla^2 V = 0 \qquad (1)$$

Because of the linearity of the $\nabla^2$ operator, the potential $V$ can be expressed as a linear combination of the basic solutions of Laplace's equation, e.g. using the spherical harmonic functions $Y_{lm}(\theta,\varphi)$:

$$V(\mathbf{r}) = \sum_{l=0}^{\infty}\sum_{m=-l}^{l} \alpha_{lm}\frac{Y_{lm}(\theta,\varphi)}{r^{l+1}} + \sum_{l=0}^{\infty}\sum_{m=-l}^{l} \beta_{lm} r^l Y_{lm}(\theta,\varphi). \qquad (2)$$

where $\alpha_{lm}$ and $\beta_{lm}$ are scalars and $r = \|\mathbf{r}\|$.

The first part (A) of the expansion diverges at the origin, thus characterizing sources that are closer to the center of the expansion than any of the sensors. In contrast, the second part (B) diverges at infinity, and corresponds to sources that are more distant to the center of the expansion than any of the sensors.

The magnetic subspace is formed by calculating the signal vectors $\mathbf{a}_{lm}$ and $\mathbf{b}_{lm}$ corresponding to the individual terms of the expansions A and B, respectively, up to sufficiently high orders $L_{in}$ and $L_{out}$. Then any measured signal vector $\phi$ is expressed as a linear combination of the basis vectors (here the monopole term $l = 0$ is left out):

$$\phi = \sum_{l=1}^{L_{in}} \sum_{m=-l}^{l} \alpha_{lm}\mathbf{a}_{lm} + \sum_{l=1}^{L_{out}} \sum_{m=-l}^{l} \beta_{lm}\mathbf{b}_{lm}. \quad (3)$$

leading to a compact matrix notation

$$\phi = \mathbf{Sx} = [\mathbf{S}_{in}\ \mathbf{S}_{out}] \begin{bmatrix} \mathbf{x}_{in} \\ \mathbf{x}_{out} \end{bmatrix}, \quad (4)$$

where
$$\mathbf{S}_{in} = [\mathbf{a}_{1,-1} \ldots \mathbf{a}_{L_{in}L_{in}}],$$
$$\mathbf{S}_{out} = [\mathbf{b}_{1,-1} \ldots \mathbf{b}_{L_{out}L_{out}}],$$
$$\mathbf{x}_{in} = [\alpha_{1,-1} \ldots \alpha_{L_{in}L_{in}}]^T,$$
$$\mathbf{x}_{out} = [\beta_{1,-1} \ldots \beta_{L_{out}L_{out}}]^T. \quad (5)$$

This approach offers an elegant method to construct the magnetic subspace by starting from the lowest spatial frequencies and adding basis vectors to $\mathbf{S}$ until they start to represent fine details having amplitudes that are below the noise level of the device. Furthermore, the biomagnetic signals will be spanned by the basis $\mathbf{S}_{in}$ and the external disturbances by the basis $\mathbf{S}_{out}$, if the origin of the expansions is placed inside the volume including the interesting sources.

The possibility to divide the interesting signals and interference signals into separate subspaces can be understood from the schematic illustration in figure 1. Here $\mathbf{I}_{in}$ and $\mathbf{I}_{out}$ describe the interesting and interference sources, respectively. The harmonic potentials associated with these sources are given in the different volumes by either A- or B-parts of the expansion as indicated. Specifically, in volume 3 where the sensor array is located, the potential associated with $\mathbf{I}_{in}$ is given by the A-part of the expansion, and the potential associated with $\mathbf{I}_{out}$ is given by the B-part of the expansion. The resolution between the interference and interesting magnetic subspaces in the SSS method is based on this fact.

The dimension of the SSS basis as a function of the orders $L_{in}$ and $L_{out}$ is given as

$$n = (L_{in} + 1)^2 + (L_{out} + 1)^2 - 2. \quad (6)$$

In practice, $L_{in} = 9$ is sufficient for biomagnetic sources and even for the most complicated interference $L_{out} = 6$ is enough. Consequently, $n = 147$ justifying the applicability of the SSS method for modern multichannel measurement devices as the fundamental requirement is $N > n$, where $N$ is the number of channels.

With this requirement fulfilled, one gets a linearly independent SSS basis spanning the subspace of all measurable magnetic signals. In this basis, the interesting signals and the interference signals are simultaneously modeled with separate signal vectors. Because the SSS basis is linearly independent, this separation is unique, and the interesting signal can easily be reconstructed from the estimated components corresponding to $\mathbf{S}_{in}$:

$$\hat{\mathbf{x}} = \begin{bmatrix} \hat{\mathbf{x}}_{in} \\ \hat{\mathbf{x}}_{out} \end{bmatrix} = \mathbf{S}^\dagger \phi$$

$$\hat{\phi}_{in} = \mathbf{S}_{in}\hat{\mathbf{x}}_{in},$$

where $\mathbf{S}^\dagger$ is the pseudoinverse of $\mathbf{S}$.

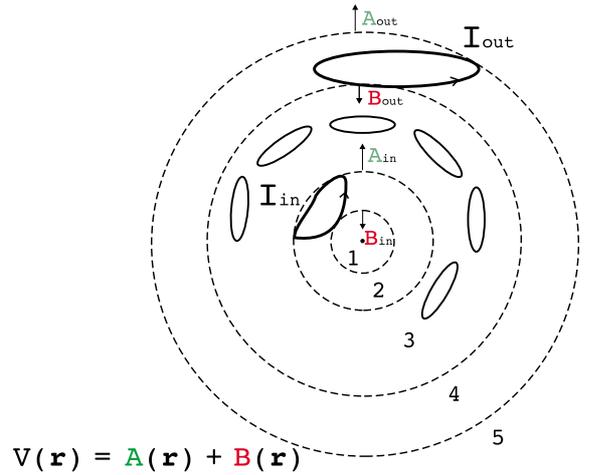

Fig. 1 Geometry of a typical neuromagnetic measurement including the interesting source and a disturbance source. The origin is in the center.

The harmonic amplitudes can be calculated in the head coordinate system. Then $\mathbf{x}_{in}$ is device-independent and can be used to transform biomagnetic signals between different sensor arrays. The transformation is done simply by using the basis $\mathbf{S}_{in}$ of the desired sensor array in the SSS reconstruction, and this array need not be the same that was used for measuring the signal and determination of the harmonic amplitudes. The same idea generalizes to a movement correction method, provided that a continuous movement detection is available. In this correction method, one calculates and possibly averages the harmonic amplitudes attached to the head and calculates the signals in a virtual array locked to the subject's head. The basic idea is the same as that described in [5] extended by the ability to remove the external disturbances: the components of $\mathbf{x}$ are estimated from the measured signal vector and then $\mathbf{x}_{in}$ is used in reconstructing the signal corresponding to a reference head position by $\hat{\phi}_0 = \mathbf{S}_{0,in}\hat{\mathbf{x}}_{in}$, where $\mathbf{S}_{0,in}$ corresponds to the reference head position.

A very interesting application of the above movement correction method is its possibility to perform DC measurements. The SQUID sensors used in MEG devices are insensitive to static fields. However, when the subject moves, the DC sources produce time-varying signals detected by the SQUIDs. When performing movement correction by always estimating the harmonic amplitudes in the head coordinate system, the time-varying signal caused by the DC sources modulated by the movement will be demodulated and appears as a static component in $x_{in}$. In this way, movement-related artifacts caused by magnetic impurities, for example, can be trivially removed from this movement-corrected result by a baseline correction. On the other hand, interesting biomagnetic DC currents can be measured using voluntary head movements followed by a movement correction as described above. This leads to much easier DC measurements than proposed earlier [6].

SSS relies on the physics and geometry of the magnetic fields and the sensor array. Thus, it is of utmost importance to know the calibration and geometry of the measurement device as accurately as possible. The achievable shielding factor against external interference is roughly the inverse of the relative calibration accuracy. An accuracy better than 1 % is crucial for SSS to perform optimally.

## Results

Figures 2 and 3 demonstrate the SSS reconstruction in the case of an evoked response of a newborn measurement. The response is expected to be seen on the occipital area of the sensor array but no dipolar field pattern can be recognized from the original data which is dominated by interference signals arising from sources both far away and in the immediate vicinity of the helmet. The interference contaminates the magnetometer signals badly but it also affects the gradiometers as can be seen in figure 3. Thus, the relatively small but significant artifact on the gradiometers could be interpreted as a brain response if no interference reduction were done.

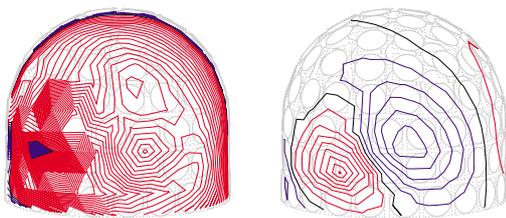

Fig. 2 The field distribution of a newborn measurement before (left) and after (right) the SSS reconstruction. Both figures are based on magnetometer signals and have the same contour step.

Figure 3 shows a close-up of the signals of some of the occipital channels. The magnetometers show a large-amplitude low frequency drift superimposed on a fast artifact generated in the vicinity of the sensor array. Being a spatially complex field, the nearby artifact affects some of the gradiometers also. All artifacts are removed by the SSS reconstruction.

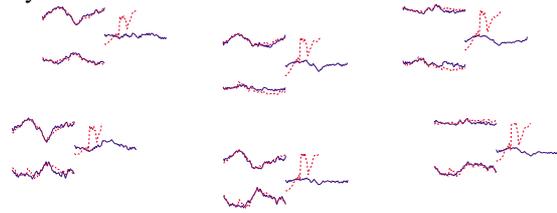

Fig. 3 Signals of some of the occipital channels before (dotted red) and after (solid blue) the SSS reconstruction.

Figure 4 demonstrates that SSS does not mix signals coming from external and internal sources. The original empty room signal of the magnetometer channel is dominated by low frequency fluctuations and the 50 Hz power line interference. In the absence of internal sources the SSS reconstructed signal $b_{in}$ is practically zero while there is a very high correlation between the original signal and the SSS reconstructed $b_{out}$. The rejection ratio against the external disturbances is seen to be several hundreds.

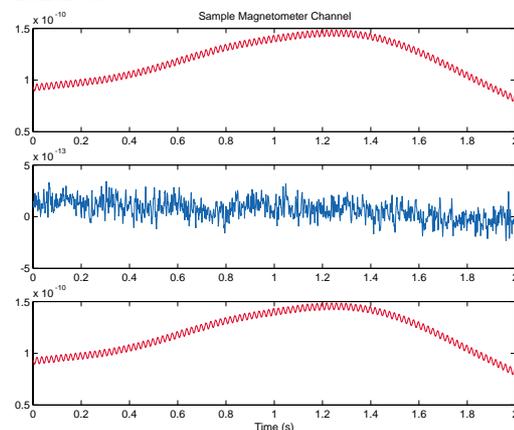

Fig. 4 Upper figure shows the original empty room signal b. The second and third figures are the SSS reconstructed interesting $b_{in}$ and disturbance $b_{out}$ respectively. Note that the figure showing $b_{in}$ is a close-up with the amplitude scale by factor 100 smaller than in the other figures.

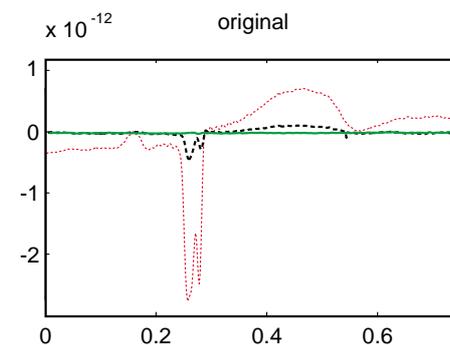

Fig. 5 Removing baby heart artifact. Dotted red; as measured, dashed black; field derivatives up to order 2 removed, solid green; field derivatives up to order 5 removed.

Figure 5 illustrates the complexity of an interference signal generated by a source in the immediate vicinity of the sensor array: the heart of a newborn subject. In this case a sufficient artifact rejection requires an interference subspace including derivates up to order 5, corresponding to $L_{\text{out}} = 6$.

The movement correction along with the idea to do DC measurements was tested by performing a SEF measurement for a subject with small magnetic particles attached to the surface of the head. First, a reference SEF measurement was performed while the subject stayed absolutely still. After that, the subject voluntarily moved his head continuously. The movement made the static fields of the small magnetized particles appear as time-varying signals in the sensor output. Consequently, the averaged response was contaminated by the movement artifact. However, by using the information obtained by the continuous head position monitoring for demodulating the AC signals due to the magnetized particles, the SEF response was recovered, as illustrated in figure 6.

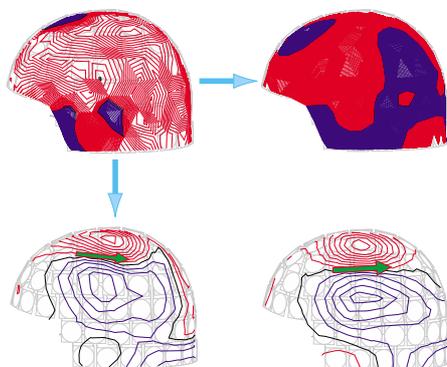

Fig. 6 Somatosensory recording with magnetic impurities on the subject's head. Upper left; data as recorded with head moving. Lower left; SEF signal after impurity artifact removed by SSS. Upper right; the demodulated DC field. Lower right; reference SEF recording with head immobilized.

## Discussion

SSS is a new method to remove external disturbances and movement artifacts, to calculate virtual signals, and to perform movement correction and DC measurements. SSS efficiently exploits the fact that the number of channels in modern multichannel MEG devices clearly exceeds the number of degrees of freedom of the measurable magnetic fields.

By using harmonic functions, SSS creates a fundamental subspace, the magnetic subspace, for all measurable multichannel signals of magnetic origin. Furthermore, SSS models both the interesting and external interference signals simultaneously by separate, linearly independent subspaces for the signals caused by sources inside and outside of the sensor array, respectively. Consequently, any measured signal can be uniquely decomposed into separate components representing the biomagnetic signals arising from inside of the array and external interference signals arising from outside of the array.

In addition to the interference removal, the device-independency of the harmonic amplitudes calculated in the signal decomposition leads to further applications for the SSS method. The most obvious of these is the transformation of the measured signals into any desired virtual sensor array. The same idea can be used for a robust movement correction method in which the movement of the head is taken into account in calculating the device-independent harmonic amplitudes that are used in transforming the signals to a sensor array locked to the subject's head. It also turns out that the movement correction method automatically enables one to do DC measurements. This is based on the fact that the DC signals are measurable because the movement-induced signals will appear as static components in the harmonic amplitudes related to the head coordinate system.

As a conclusion, SSS greatly improves the quality of MEG data without requiring extensive user intervention, a particularly important feature in clinical MEG work. SSS allows measurements with varying interference sources in the environment, even inside the shielded room and on the subjects. Even moving subjects having moderately magnetic impurities or implants can be measured and the data analyzed.

## Acknowledgements

The authors would like to thank prof. Patricia Kuhl, prof. Toshiaki Imada, and prof. Marie Cheour for providing us with their infant MEG data that formed a crucial stimulation for the SSS method development.